\documentclass[prl, aps, twocolumn, amsmath, amssymb, 10pt]{revtex4-2}

\usepackage{amsmath,graphicx,color}
\usepackage{bm}
\usepackage{natbib}
\usepackage{placeins}
\setcitestyle{square}
\usepackage{lipsum}
\usepackage{braket}
\usepackage{leftindex}
\usepackage{makerobust}
\usepackage{xcolor}
\usepackage{hyperref}
\hypersetup{%
 colorlinks,
 breaklinks=true,
 plainpages=false,%
 citecolor=blue,
 linkcolor=blue,
 urlcolor=blue,
 bookmarksopen=true,%
 bookmarksnumbered=false,%
 bookmarksdepth=5%
}

\newcommand{\makeauthor}[2]{\newcommand{#1}[1]{{%
 \sffamily\color{#2}{%
 \bfseries\begingroup\escapechar=-1\edef\x{\endgroup\string#1}\x:%
 } ##1}}%
 \MakeRobustCommand#1}
\makeauthor{\dkm}{red}
\makeauthor{\rew}{blue}
\makeauthor{\jb}{purple}

\begin{document}

\title{Majorana Edge Modes as Quantum Memory for Topological Quantum Computing} 

\author{Jasmin Bedow and Dirk K. Morr \\
{\textit{Department of Physics, University of Illinois Chicago, Chicago, IL 60607, USA}} \\
\today
}

\begin{abstract}
We demonstrate that a combination of Majorana edge modes (MEMs) and Majorana zero modes (MZMs) located in the vortex cores of two-dimensional topological superconductors represent a new platform for the efficient implementation of fault-tolerant quantum gates. By calculating the full many-body dynamics of the system, we demonstrate the successful simulation and visualization of $Z$-, $X$- and Hadamard gates, with MEMs being functionalized as quantum memory. Our results open a new platform for the efficient implementation of fault-tolerant quantum computing.
\end{abstract}

\maketitle

Topological superconducting phases in two-dimensional (2D) systems host low-energy Majorana edge modes (MEMs) on the boundaries of topological domains \cite{Menard2017, Palacio-Morales2019, Kezilebieke2020, Bazarnik2023, Soldini2023, Bruening2024, LoConte2025} and Majorana zero modes (MZMs) in defects such as magnetic vortices \cite{Menard2019,LoConte2025,Wang2018,Machida2019,Kong2019,Zhu2020}. While many proposals have been made to utilize MZMs in 1D structures for the implementation of fault-tolerant quantum computing \cite{Nayak2008}, requiring an atomic scale control of the system's magnetic and electronic properties \cite{Alicea2011,Halperin2012,Sekania2017,Harper2019,Tutschku2020,Tanaka2022,Amorim2015,Kraus2013,Aasen2016,Karzig2017,Zhou2022,Hyart2013,Li2016,Sanno2021,Cheng2016,Chen2022,Wong2023}, MZMs localized in magnetic vortices offer the advantage that they can in principle be braided by dragging the vortices using scanning tunneling microscopy (STM) \cite{Ge2016}, magnetic force microscopy \cite{Straver2008,November2019,Polshyn2019} or a superconducting quantum interference device (SQUID) techniques \cite{Gardner2002,Kalisky2011,Kremen2016}. At the same time, utilizing MEMs for quantum gates has received only limited attention \cite{Beenakker2019_2,Flor2023}, raising the intriguing question of whether a combination of MEMs and MZMs could be employed as a new platform for the realization of fault-tolerant quantum computing.

\begin{figure}[t]
 \centering
 \includegraphics[width=\columnwidth]{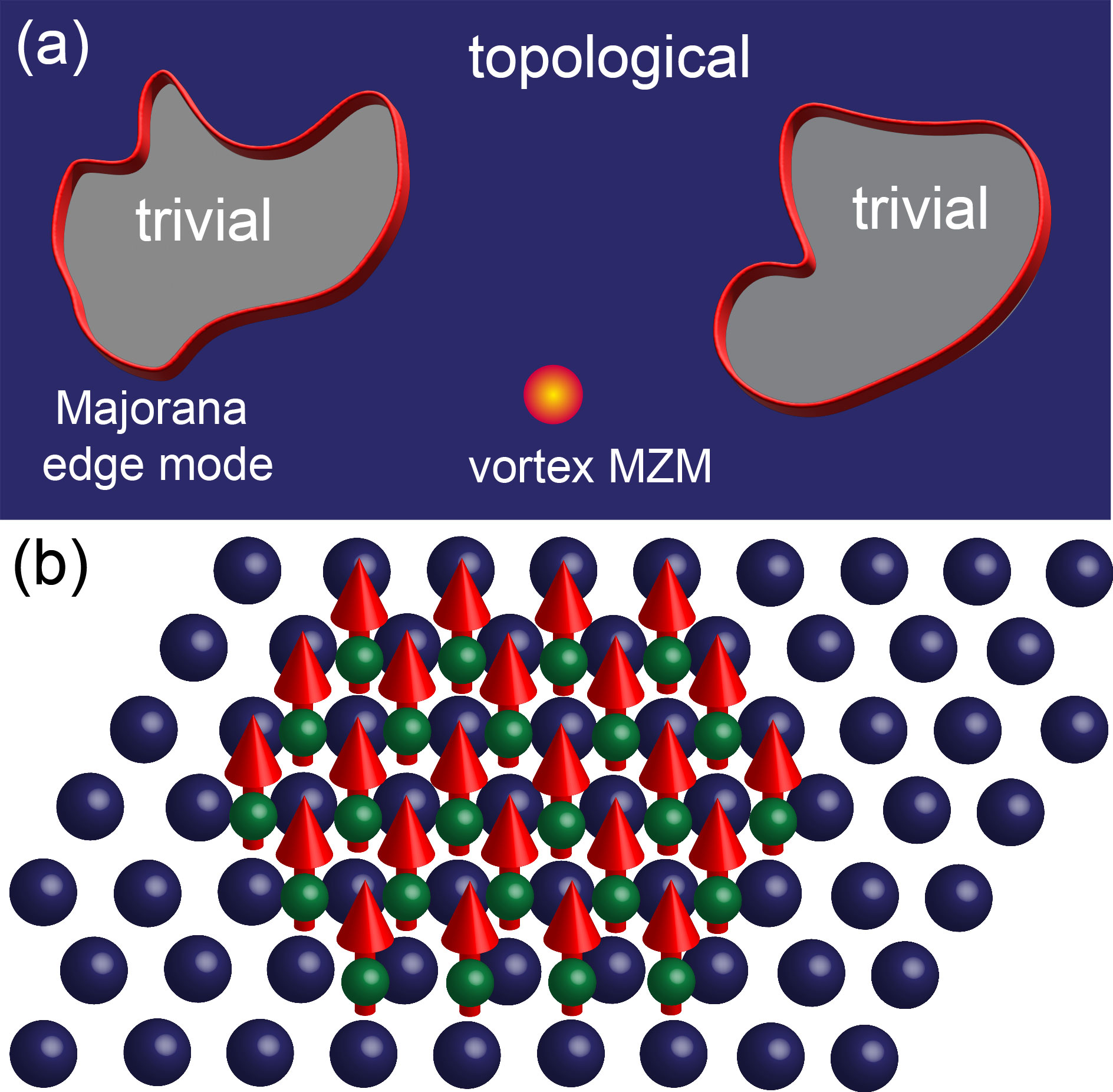}
 \caption{
    (a) Schematic of 2D topological superconductors containing topological and trivial domains possessing MEMs and vortex core MZMs. 
    (b) Domain structure in MSH systems containing magnetic islands.
 } 
 \label{fig:Fig1}
\end{figure}

In this Letter, we demonstrate how a combination of MEMs and vortex core MZMs in 2D superconductors that contain both topological and trivial domains can be employed to successfully realize topologically protected $Z$-, $X$- and Hadamard gates. Such a domain structure was proposed to occur due to disorder in FeSe$_{1-x}$Te$_{x}$ [Fig.~\ref{fig:Fig1}(a)] \cite{Mascot2022, Xu2023} and can also be realized in magnet-superconductor hybrid (MSH) systems containing magnetic islands \cite{Menard2017, Palacio-Morales2019, Kezilebieke2020} [Fig.~\ref{fig:Fig1}(b)]. In particular, we show that while qubit states can be stored in MEMs -- rendering them an effective quantum memory -- vortices can be employed to execute quantum gates on these states. This process is achieved by moving vortices through the edges of topological phases, effectively converting a delocalized MEM into an MZM localized in the vortex core. Using recent advances in simulating non-equilibrium processes in superconductors \cite{Mascot2023,Bedow2024,Bedow2025}, we compute the full many-body dynamics of these processes on time scales from a few fs to ns to analyze the quantum gates' success probabilities. Furthermore, we visualize gate processes using the non-equilibrium local density of states (LDOS), which is proportional to the time-dependent differential conductance measured in scanning tunneling microscopy (STM) experiments \cite{Bedow2022}. Our results thus establish a new material platform for the realization of fault-tolerant topological quantum computing. 

{\bf Theoretical Methods.~} To study a 2D superconductor containing both topological and trivial domains in the presence of magnetic vortices, we consider a system in which topological superconductivity arises from the interplay of $s$-wave superconductivity, Rashba spin-orbit coupling and ferromagnetism \cite{Ron2015,Li2016,Rachel2017}, as described by the Hamiltonian
\begin{align}
\mathcal{H} =& \; \sum_{{\bf r}, {\bf r}', \beta, \gamma} e^{\mathrm{i} \theta({\bf r}, {\bf r}')} c^\dagger_{{\bf r}, \beta} \left[ -t_e \delta_{\beta \gamma} + \mathrm{i} \alpha \left(\hat{e}_{{\bf r}' - {\bf r}} \times \boldsymbol{\sigma} \right)^z_{\beta\gamma} \right] c_{{\bf r}', \gamma} \nonumber \\
&- \mu \sum_{{\bf r}, \beta} c^\dagger_{{\bf r}, \beta} c_{{\bf r}, \beta}
 + \sum_{{\bf r}} |\Delta_{\bf r}(t)| \left( e^{\mathrm{i} \phi({\bf r}, t)} c^\dagger_{{\bf r}, \uparrow} c^\dagger_{{\bf r}, \downarrow} +h.c. \right) \nonumber \\ 
&- J{\sum_{{\bf R},\beta}}^\prime c^\dagger_{{\bf R}, \beta} \left[{\bf S}_{\bf R} \cdot {\bm \sigma}\right]_{\beta\gamma} c_{{\bf R}, \gamma} \; . \label{eq:ham}
\end{align}
Here, $c^\dagger_{{\bf r}, \beta}$ creates an electron with spin $\beta$ at site ${\bf r}$, $-t_e$ denotes the nearest-neighbor hopping parameter on a 2D square lattice, $\alpha$ is the Rashba spin-orbit coupling between nearest-neighbor sites ${\bf r}$ and ${\bf r}'$, $\mu$ is the chemical potential, $|\Delta_{\bf r}|$ is the magnitude of the $s$-wave superconducting order parameter (SCOP) at site ${\bf r}$, and $J$ is the magnetic exchange coupling between the magnetic adatom with spin ${\bf S}_{\bf R}$ of magnitude $S$ at site ${\bf R}$ and the conduction electrons. As the hard superconducting gap suppresses Kondo screening \cite{Balatsky2006,Heinrich2018}, we consider the magnetic adatoms as classical spins. Below, we model the topological domain for concreteness as a region that is covered by magnetic adatoms, while the trivial domain is not (the primed sum then runs over all sites of the topological domain). However, as previously proposed \cite{Xu2023}, the different nature of these domains could also arise from spatial variations in the chemical potential, which we expect to yield similar results to the ones discussed below. A ferromagnetic out-of-plane alignment of the magnetic moments then yields a topological phase with Chern number $C=-1$ \cite{Rachel2017}. Vortices are implemented through the Peierls phase 
$\theta({\bf r}, {\bf r}')= \frac{\pi}{\phi_0} \int_{\bf r}^{\bf r'} {\bf A} \cdot \mathrm{d}{\bf l}$ arising from the vector potential ${\bf A}$ (for details, see Supplemental Material (SM) Sec.~I). The vortex motion is encoded in a time-dependent magnitude $|\Delta_{\bf r}(t)|$ and phase $\phi({\bf r}, t)$ of the SCOP (for details, see Appendix A).  Below all times are given in units of $\tau_e = \hbar/t_e$ such that for typical values of $t_e$ of a few hundred meV, $\tau_e$ is of the order of a few femtoseconds. The characteristic time for the vortex motion, $t_V$, is chosen to be much larger than $\hbar \:/\: \Delta_t$, where $\Delta_t$ is the topological gap, such that any excitations between the topological MZM states and the trivial Caroli-de Gennes Matricon (CdGM) states inside the vortex core are avoided \cite{Caroli1964} (for details, see Appendix A). To evaluate the success of the gate operations, we define transition probabilities between a time-evolved many-body state $\ket{m(t)}$ at time $t$ and a state $\ket{n}_i$ $(i=E,C)$ in the MEM occupation space ($i=E$), or in the combined occupation space of MEMs and MZMs ($i=C$). The initial state $\ket{m(t=0)} = \ket{m}_E$ is always defined for the spatial configuration prior to the start of the gate process with vortices located in the trivial region, while $\ket{n}_C$ is a static reference state, as specified below, when the vortices are in the topological region (for details, see SM Sec.~II and Refs.~\cite{Mascot2023,Bedow2024}).
To gain insight into the exchange statistics of MZMs and MEMs, revealed by a $\sqrt{Z}$-gate, we compute the gauge- and parametrization-invariant geometric phase difference $\Delta \phi$ between the even- and odd-parity states \cite{Samuel1988, Mukunda1993,Bedow2024} (for details, see Appendix B and SM Sec.~II). The Ising anyon statistics of MEMs or MZMs then results in $\Delta \phi$ being an odd integer multiple of $\pi/2$. Finally, we visualize the entire gate process and motion of vortices using the energy-, time- and spatially-resolved non-equilibrium density of states $N_\mathrm{neq}({\bf r}, \sigma, t, \omega)$ \cite{Bedow2024,Bedow2025}, 
which is proportional \cite{Bedow2022} to the time-dependent differential conductance $\mathrm{d}I(V,{\bf r},t)/\mathrm{d}V$ measured in STM experiments. 

{\bf Results.~}
{\it $\sqrt{Z}$- and $Z$-gate of Majorana Edge Modes} 
\begin{figure}[ht]
 \centering
 \includegraphics[width=\columnwidth]{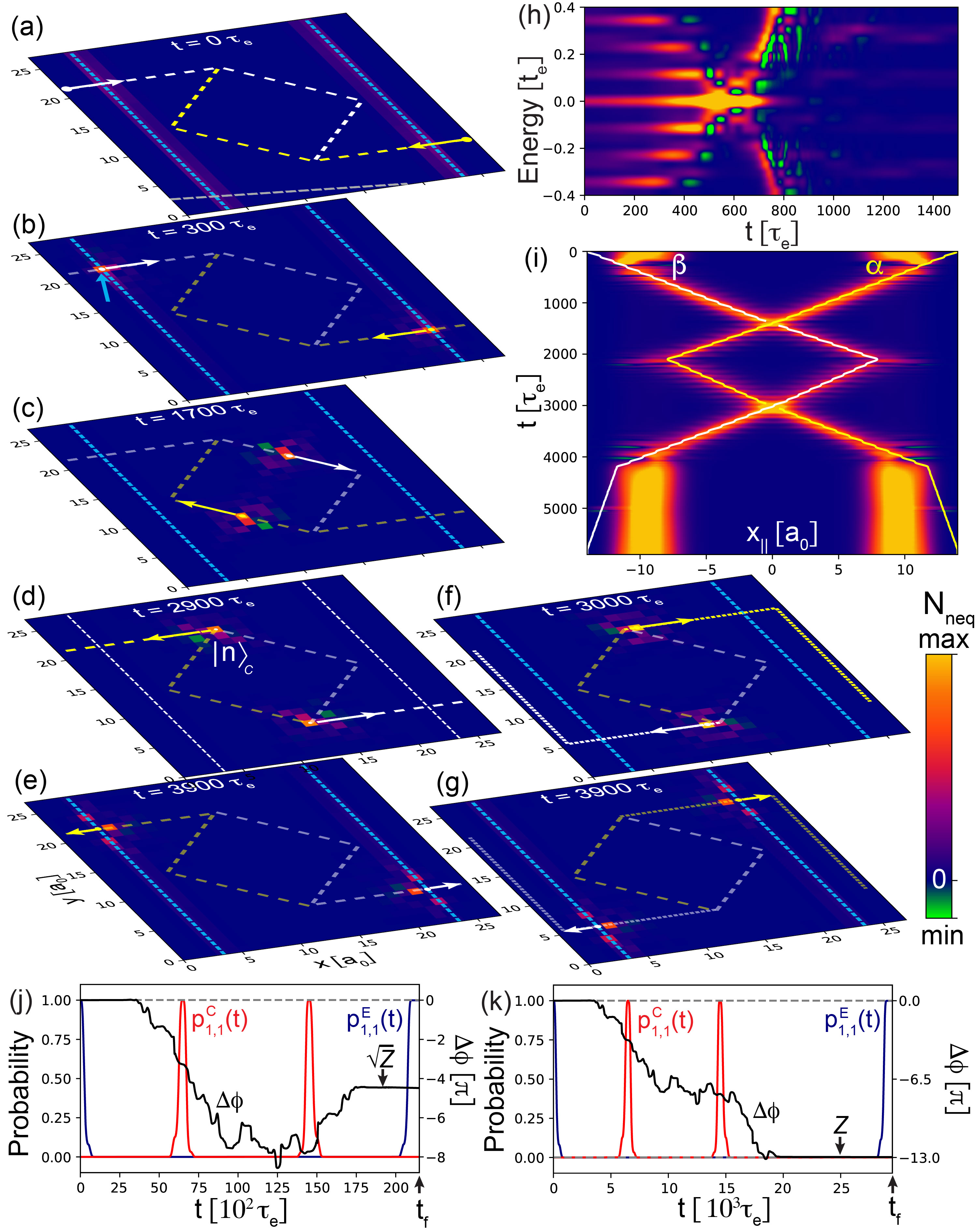}
 \caption{
 Zero-energy $N_\mathrm{neq}$ for various times during a (a)-(e) $\sqrt{Z}$- and (a)-(c),(f),(g) $Z$-gate process. (h) Energy- and time-resolved $N_\mathrm{neq}$ at the domain wall site that the vortex passes through at $t=300\tau_e$ (marked by a blue arrow in (b)). (i) Majorana world lines for the $Z$-gate process as a function of time, obtained by projecting $N_\mathrm{neq}$ onto the grey axis in (a). 
 Time-dependent transition probabilities $p_{1,1}^{C,E}(t)$ 
 and geometric phase difference $\Delta \phi(t)$ (black) for the (j) $\sqrt{Z}$- and (k) $Z$-gate. Parameters are $(\mu, \alpha, \Delta, JS, \Gamma) = (-4, 0.9, 2.4, 5.2,0.01) t_e$ with $R_V = 3$, $t_V = 100 \tau_e$ in (a)-(g),(i), and $R_V = 1$, $t_V = 200 \tau_e$ in (h), $t_V = 500 \tau_e$ in (j),(k).
 }
 \label{fig:Fig2}
\end{figure}
We begin by studying the Ising anyon statistics of MEMs and to this end consider a system with trivial and topological domains in the form of a ribbon geometry with periodic boundary conditions, as shown in Fig.~\ref{fig:Fig2}(a). Here, the MEMs are delocalized along the domain walls (indicated by dashed blue lines) and two vortices (indicated by white and yellow dots) are initially located in the trivial domain. To define qubit states in terms of MEMs, we need to create pairs of MEMs with the MEMs located on  opposite or the same domain walls \cite{Bedow2024}. While $\sqrt{Z}$-gates can be simulated with either choice, we begin by considering the first case, creating a pair of zero-energy MEMs from opposite edges, which can be achieved, for example, by using an even number of sites $N_y$ in the $y$-direction (the second case will be discussed below). The zero-energy $N_\mathrm{neq}$ for various times during the $\sqrt{Z}$- and $Z$-gate processes are shown in Figs.~\ref{fig:Fig2}(a)-(g). With the vortices initially located in the trivial region, the two zero-energy MEMs lead to a finite LDOS along the domain walls, as evidenced by the zero-energy $N_\mathrm{neq}$ shown in Fig.~\ref{fig:Fig2}(a). When the vortices are moved into the topological region through the domain walls, the two delocalized zero-energy MEMs are transferred to the vortex cores as localized MZMs [Fig.~\ref{fig:Fig2}(b)]. This becomes particularly apparent when plotting the energy- and time-dependent $N_\mathrm{neq}$ at the domain wall site through which a vortex moves, demonstrating the "vanishing" of the zero-energy MEM from the domain wall as the vortex enters the topological domain [see Fig.~\ref{fig:Fig2}(h)]. The two vortices are subsequently exchanged, following the exchange path denoted by a dashed line in Fig.~\ref{fig:Fig2}(c). After this exchange, the vortices can be moved back into the trivial domain either by returning them via the same domain wall they entered through [see Figs.~\ref{fig:Fig2}(f) and (g)], or via the opposite one [see Figs.~\ref{fig:Fig2}(d) and (e)]. In the former case, the MEMs are exchanged twice via the vortex core MZMs, realizing a $Z$-gate, as reflected in the Majorana world lines shown in Fig.~\ref{fig:Fig2}(i) [obtained by projecting $N_\mathrm{neq}$ onto the gray axis in Fig.~\ref{fig:Fig2}(a)] while in the latter case they are exchanged only once, realizing a $\sqrt{Z}$-gate (the full time dependence of $N_\mathrm{neq}$ for the $\sqrt{Z}$- and $Z$-gate processes is shown in Supplementary Movies 1 and 2, respectively). 
This is confirmed by a plot of the geometric phase difference, $\Delta \phi$, which reveals that $\Delta \phi=-9\pi/2$ at the end of the $\sqrt{Z}$-gate process [Fig.~\ref{fig:Fig2}(j)], while $\Delta \phi = -13\pi$ at the termination of the $Z$-gate process [Fig.~\ref{fig:Fig2}(k)], thus confirming the Ising anyon statistics of the MEMs. The required adiabaticity of these processes is reflected in $p^E_{1,1}(t=t_f)$ reaching unity at the end of the gate process at $t=t_f$ [see Figs.~\ref{fig:Fig2}(j) and (k)]. In addition, defining a topological reference state $\ket{1}_C$ as the one when the two vortex core MZMs are located in the spatial positions shown in Figs.~\ref{fig:Fig2}(d), we find that the initial many-body wave function, $\ket{1(t=0)}=\ket{1}_E$, defined in terms of two MEMs, evolves adiabatically into the many-body wave function $\ket{1}_C$ defined in terms of the two MZMs, as reflected in $p^C_{1,1}(t)$ reaching unity at two intermediate times during the gate process [see Figs.~\ref{fig:Fig2}(j),(k)]. This result opens the intriguing possibility to utilize MEMs to initialize the many-body states of vortex core MZMs.

\begin{figure}[ht]
 \centering
 \includegraphics[width=\columnwidth]{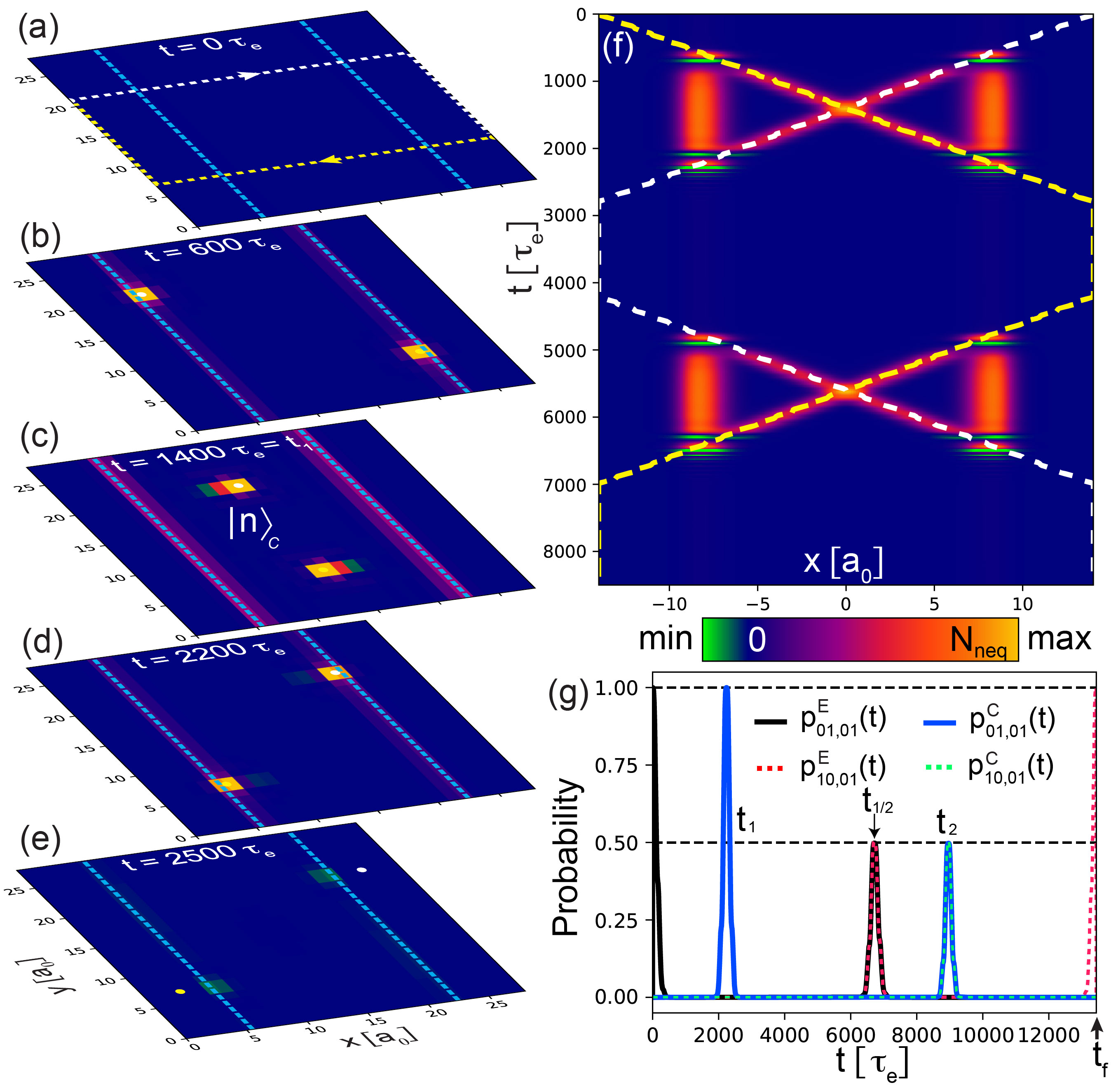}
 \caption{(a)-(e) Zero-energy $N_\mathrm{neq}$ for various times during an $X$-gate process.
 (f) Majorana world lines for the $X$-gate process as a function of time, obtained by projecting the zero-energy $N_\mathrm{neq}$ onto the $x$-axis.
 (g) Time-dependent transition probabilities $p_{n,01}^{C,E}(t)$ . 
 Parameters are $(\mu, \alpha, \Delta, JS, \Gamma) = (-7, 0.9, 2.4, 4.4, 0.01) t_e$, $R_V = 1$ with (a),(b) $t_V = 100 \tau_e$ and (c),(d) $t_V = 160 \tau_e$. 
 }
 \label{fig:Fig3}
\end{figure}

{\it Non-Abelian Braiding Statistics of Majorana Edge Modes.~}
To demonstrate the non-Abelian braiding statistics of MEMs, we consider the effects of an $X$-gate. Using the same system geometry as in Fig.~\ref{fig:Fig2}, we now utilize 
pairs of MEMs from the same domain wall to define the topological qubit states. Each domain wall thus provides a pair of degenerate low-energy MEMs with energy $\pm \epsilon_0$ (this pairing can be realized, for example, by choosing $N_y$ to be odd), yielding a total of 4 MEMs necessary to define an $N=1$ qubit state. To execute an $X$-gate on these states, an MEM from one pair located at one domain wall needs to be exchanged twice with an MEM from the opposite domain wall; an exchange that can be facilitated by moving two vortices along the closed loop shown in Fig.~\ref{fig:Fig3}(a) as a dashed line.

In Figs.~\ref{fig:Fig3}(a)-(e), we present the zero-energy $N_\mathrm{neq}$ for various times during the first half of $X$-gate process (the full time dependence of this gate is shown in Supplementary Movie 3). As a vortex moves through a domain wall, one of the MEMs with energy $\pm \epsilon_0$ is transferred to the vortex as an MZM, while its MEM partner remains localized at the domain wall [Fig.~\ref{fig:Fig3}(b)]. With increasing distance between the vortex and the domain wall, the energy of these two states approaches zero due to a vanishing hybridization, resulting in a considerable increase in spectral weight at zero energy in the vortex core as well as along the domain wall [Fig.~\ref{fig:Fig3}(c)]. The double exchange of MEMs via vortex core MZMs is clearly seen in the Majorana world lines shown in Figs.~\ref{fig:Fig3}(f), obtained by projecting the zero-energy $N_\mathrm{neq}$ onto the $x$-axis (the path of the two vortices is shown by dashed white and yellow lines).
A plot of the transition probabilities in the odd-parity sector in Fig.~\ref{fig:Fig3}(g) (we obtain analogous results in the even-parity sector) shows that after the vortices have moved halfway around the path in Fig.~\ref{fig:Fig3}(a) at time $t=t_{1/2}$, the initial $\ket{01}_E$ state has evolved into an equal superposition of $\ket{01}_E$ and $\ket{10}_E$ state, i.e. into the state $\ket{+}_E=\alpha \ket{01}_E+\beta \ket{10}_E$, as evidenced by $p^E_{01,01}(t_{1/2})=p^E_{01,10}(t_{1/2})=1/2=|\alpha|^2=|\beta|^2$, thus realizing a $\sqrt{X}$-gate. The MZMs carried by the vortices across the topological domain thus facilitate a change in the topological qubit state of the system which, after the vortices enter the trivial region, is stored in the MEMs, which are functionalized as quantum memory. After the second half of the vortex path is completed at $t=t_f$, we obtain $p^E_{10,01}(t_f)=0.999$, which is above the threshold for quantum error correction \cite{Raussendorf2007,Straver2008} and thus represents a successful simulation of an $X$-gate using MEMs, demonstrating their non-Abelian braiding statistics.

We note that while the two vortices are in the topological domain, the $N=1$ qubit state, originally defined in terms of the MEM occupation space, is now encoded in the space of the two vortex core MZMs and their two MEM partners, which we refer to as a mixed-encoded qubit state. To demonstrate this we employ the spatial position of the vortices shown in Fig.~\ref{fig:Fig3}(c) (when they are in the middle of the topological domain), to define mixed-encoded qubit states $\ket{10}_C$ and $\ket{01}_C$. We first note that the initial state, $\ket{01}_E$, adiabatically evolves into the mixed-encoded qubit state $\ket{10}_C$ at $t_1$, as evidenced by $p_{01,01}^C(t_1)=1$. Moreover, during the second half of the vortex path, the MEM state $\ket{+}_E$ adiabatically evolves into a superposition of the mixed-encoded $\ket{10}_C$ and $\ket{01}_C$ states at $t_2$, as follows from $p^C_{01,01}(t_{2})=p^C_{01,10}(t_{2})=1/2$, implying that even superpositions of pure MEM qubit states can be transferred to a mixed-encoded state. This platform therefore provides the unique ability to define qubit states using both MEMs and MZMs, and to adiabatically evolve pure and mixed-encoded states into each other. 

\begin{figure}[ht]
 \centering
 \includegraphics[width=\columnwidth]{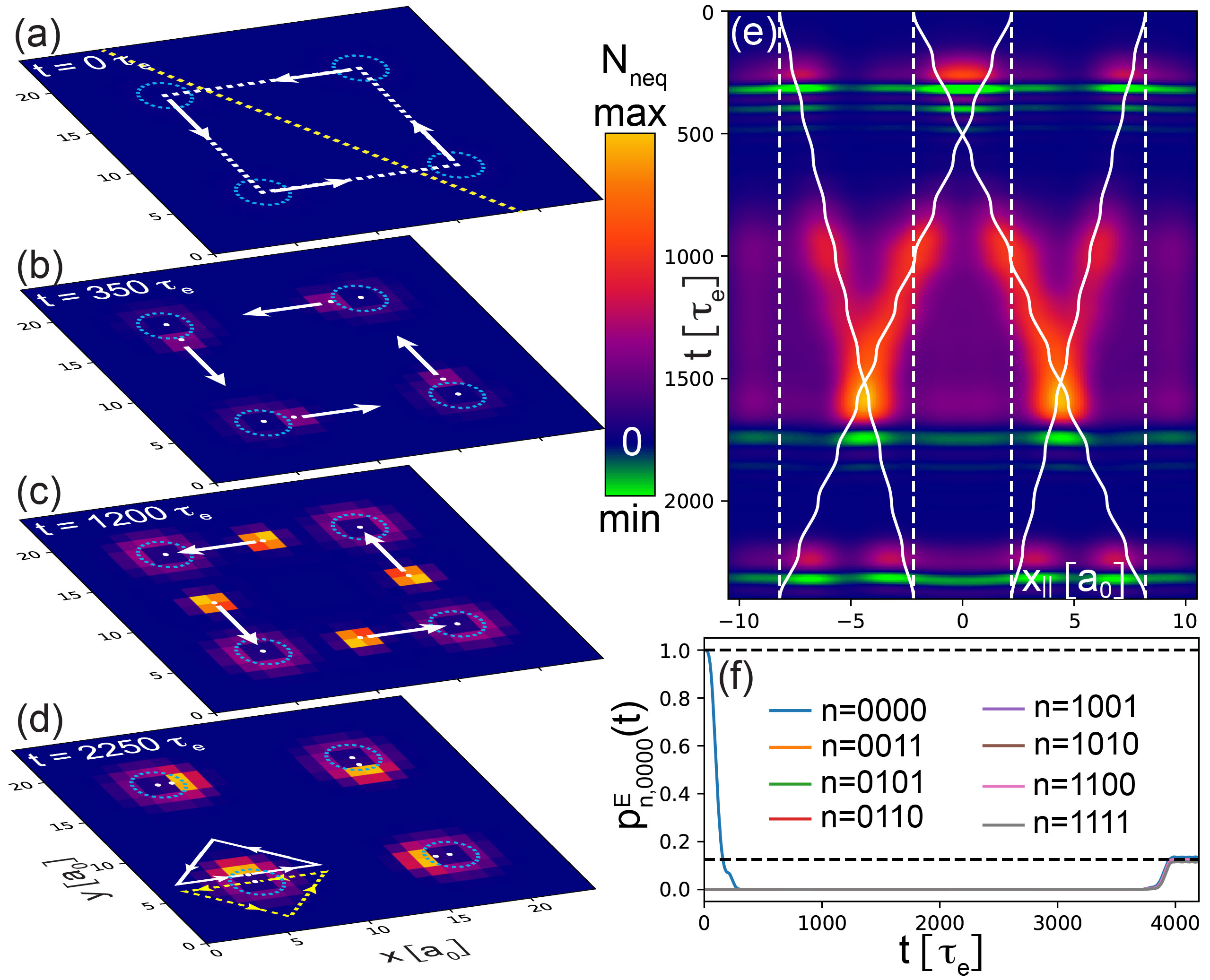}
 \caption{
 (a)-(d) Zero-energy $N_\mathrm{neq}$ for various times during a 3-qubit $\sqrt{X}$-gate process.
 (e) Majorana world lines obtained from projecting the zero-energy $N_\mathrm{neq}$ onto the dashed yellow line in (a). Solid (dashed) white lines denote the moving (stationary) vortices. 
 (f) Transition probabilities $p_{n,0000}^E(t)$ from the initial $\ket{0000}_E$ state into the 8 possible even-parity states $\ket{n}_E$.
 Parameters are $\mu, \alpha, \Delta, JS, \Gamma = (-7, 0.9, 2.4, 4.4, 0.01) t_e$ with $R_V = 1$. $t_V = 200 \tau_e$ in (a),(b) in a $24\times24$ system, and (c) in a $40\times40$ system.
 } 
 \label{fig:Fig4}
\end{figure}

{\it Extension to $N$ qubits.~}
The above demonstrated ability to define mixed-encoded qubit states and perform quantum gates using MEMs and MZMs opens new possibilities for the efficient implementation of quantum algorithms for larger $N$ qubit systems in 2D. To demonstrate this, we consider an $N=3$ qubit system, in which four trivial domain bubbles, each containing two vortices, are surrounded by a common topological region, allowing us to create multiple spatially separated MEMs within one topological region, as shown in Fig.~\ref{fig:Fig4}(a) (while the two vortices can be located anywhere in the trivial domain bubble, for computational reasons, we choose identical initial positions). To execute an $N=3$ qubit $\sqrt{X}$-gate, which, as we previously showed \cite{Bedow2025}, can be utilized as a Hadamard gate, we move one vortex from each trivial domain bubble to the next going counter-clockwise, as shown by the time-evolution of the zero-energy $N_\mathrm{neq}$ in Figs.~\ref{fig:Fig4}(a)-(d) (the full time dependence of this gate is shown in Supplementary Movie 4). 
By projecting the zero-energy $N_\mathrm{neq}$ onto the dashed yellow axis in Fig.~\ref{fig:Fig4}(a), we obtain the Majorana world lines shown in Fig.~\ref{fig:Fig4}(e). Here the dashed (solid) white lines represent the stationary (moving) vortices. As the vortices are moved through the domain walls, one of the two lowest-energy MEMs with energies $\pm \epsilon_0$ is transferred to the vortex as an MZM, which is then transferred to the next domain wall surrounding a trivial bubble, leading to a cyclic exchange of MEMs. The time dependent probabilities $p^E_{n,s}(t)$ shown in Fig.~\ref{fig:Fig4}(f) reveal that the initial state $\ket{s}_E=\ket{0000}_E$ adiabatically evolves into an equal superposition of the 8 possible even parity qubit states $\ket{n}_E$, with all transition probabilities being approximately $0.125 = 2^{-3}$. This cyclic exchange of MEMs via vortex core MZMs thus represents a realization of an $N=3$ $\sqrt{X}$-gate. Moreover, exchanging the two vortices from the same bubble using the path shown in Fig.~\ref{fig:Fig4}(d) results in a $\sqrt{Z}$-gate (see SM Sec.~IV). Thus a 2D topological superconductor containing trivial bubble domains provides a suitable platform for the efficient implementation of MEMs as quantum memories, and execution of $\sqrt{X}$- and $\sqrt{Z}$-quantum gates using vortices, which are the basic building blocks for complex quantum algorithms \cite{Bedow2025}.

{\bf Discussion.~}
We have demonstrated that a combination of MEMs and MZMs, emerging in 2D topological superconductors containing vortices in trivial and topological regions, is ideally suited to execute topologically protected quantum gates. In particular, we showed that the MEMs function as an effective quantum memory, while the vortex core MZMs execute the gate operations. Moreover, qubit states defined in terms of MEMs adiabatically evolve into those of vortex core MZMs, providing a new approach to the initialization of the latter. Finally, a heterogeneous mixture of trivial domain bubbles inside a single topological domain provides a scalable system architecture, as the addition of each trivial domain containing two vortices increases the number of qubits by one. Investigating the implementation of quantum algorithms in such an architecture will be of great future interest. \\

{\bf Acknowledgments. ~}
We would like to acknowledge stimulating discussions with M. Buss, P. Gupta and J. Hoffman. This work was supported by the U.\ S.\ Department of Energy, Office of Science, Basic Energy Sciences, under Award No.\ DE-FG02-05ER46225.\\ 

{\bf Data Availability} \\
Original data are available at (insert link to Zenodo depository).\\

\section{A. Time-dependent Motion of Vortices}
\label{sec:vortex_motion}

To describe the movement of vortices from one lattice site to the next in a time $t_V$ (for horizontal, vertical, and diagonal directions), we utilize a smoothed-out Heaviside-function given by
\begin{equation}
    s(t, t_0, t_V) = 
    \begin{cases} 
    0 ,& t < t_0 \\
    \sin^2\left( \frac{t-t_0}{t_V}\right) ,& t_0 \leq t \leq t_0 + t_V \\
    1 ,&  t > t_0 + t_V
    \end{cases} \; .
\end{equation}
To move the $i'th$ vortex core, for example, along the $x$-direction from a site ${\bf R}_i(t_0)$ at time $t_0$ to ${\bf R}_i(t_0)+{\hat {\bf x}}a_0$ at time $t_0+t_V$, we change its position as 
\begin{equation}
    \mathbf{R}_i(t) = \mathbf{R}_i(t_0) + {\hat {\bf x}}a_0 s(t,t_0, t_V)
\end{equation}
and analogously for movements along the $y$- or diagonal directions.\\

The motion of the magnetic vortices leads to a time dependence in the magnitude $|\Delta_{\bf r}(t)|$ and phase $\phi({\bf r}, t)$ of the superconducting order parameter. The former is described by 
\begin{equation}
    |\Delta_{\bf r} (t)| = \Delta_0 \sin^2\left(\frac{\pi}{2 R_V} \min_{i=1,\cdots,N_V}\left[ \left|{\bf r} - {\bf R}_{i}(t)\right|, R_V\right]\right) \; ,
\end{equation}
whose spatial profile describes the reduction of the superconducting order parameter within a radius $R_V$ of the time-dependent location of the $i'th$ vortex core at site ${\bf R}_{i}(t)$, with $|\Delta_{\bf r} (t)| =0 $ at the center of the vortex. Here, $N_V$ is the number of vortices in a supercell. Moreover, the smoothed-out Heaviside function guarantees that the reduction of the superconducting order parameter's magnitude to zero at the vortex core occurs smoothly, which is necessary for an adiabatic evolution of the quantum gate processes. While for this spatial form, $|\Delta_{\bf r}(t)|$ decreases less steeply around the vortex center than that obtained in self-consistent calculations \cite{Smith2016}, it was previously shown that the occurring vortex core MZMs are robust with respect to the spatial profile of $|\Delta_{\bf r}(t)|$ \cite{Nagai2014}. Thus, while the spatial form of $|\Delta_{\bf r}(t)|$ utilized here leads to a larger MZM localization length in comparison to that obtained self-consistently, this does not affect the results presented in the main text.

To describe the time dependence of $\phi({\bf r}, t)$, we first determine its value when the vortices are centered at a lattice site or at the center of plaquette of four lattice sites, at time $t_0$, given by 
\begin{align}
    \phi({\bf r}, t_0) &= -\sum_{i=1}^{N_V} \sum_{p,q=-\infty}^{\infty} \arctan\left( \frac{r_y - R_{i,y}(t_0) - p N_y}{r_x - R_{i,x}(t_0) - q N_x} \right) \; ,
    \label{eq:SCOP_phase}
\end{align}
where ${\bf r}=(r_x,r_y)$ and ${\bf R}_{i}=(R_{i,x},R_{i,y})$. This form of the superconducting order parameter approximate well that obtained from the self-consistent calculations \cite{Vafek2001,Nagai2012,Smith2016} and ensures continuity of the phase across the boundary of the magnetic supercell. $N_x$ and $N_y$ are the numbers of sites in the supercell in the $x$- and $y$-directions, and $p,q$ run over all copies of the supercell due to the periodic boundary conditions. We then interpolate $\phi({\bf r}, t)$ between times $t_0$ and $t_0 + t_V$ using the smoothed-out Heaviside function 
\begin{equation}
    \phi({\bf r}, t) = \phi({\bf r}, t_0) + \left[\phi({\bf r}, t_0+t_V)-\phi({\bf r}, t_0) \right]
    s(t,t_0, t_V) \; .
\end{equation}

\section{B. Time-Dependent Physical Quantities}
\label{sec:timedep_observables}

We compute the geometric phase of a many-body state $\ket{m}$ as a function of time utilizing the gauge- and parametrization-invariant functional \cite{Samuel1988, Mukunda1993,Bedow2024}
\begin{equation}
   \begin{aligned}
       \phi_{m}(t) =& \; \mathrm{arg} \braket{m (t_0) | m (t)} - \mathrm{Im} \int_{t_0}^t \braket{m (t') | \dot{m} (t')} \mathrm{d}t' \; ,
   \end{aligned}
   \label{eq:geo-phase}
\end{equation}
The geometric phase difference $\Delta \phi = \phi_{\ket{1}} - \phi_{\ket{0}}$ between the occupied $\ket{1}$ and unoccupied $\ket{0}$ many-body states of two MEMs or MZMs then provides insights into their fractional statistics with $\Delta \phi$ being an odd multiple of $\pi/2$ for Ising anyons.\\

To visualize the entire gate process and motion of vortices in real time and space, we employ the energy-, time- and spatially-resolved non-equilibrium density of states, $N_\mathrm{neq}({\bf r}, \sigma, t, \omega) = -\frac{1}{\pi} {\rm Im} \, G^\mathrm{r}({\bf r},{\bf r}, \sigma, t, \omega)$, 
which we previously showed to be proportional \cite{Bedow2022} to the time-dependent differential conductance $\mathrm{d}I(V,{\bf r},t)/\mathrm{d}V$ measured in STM experiments. 
Therein, the time- and frequency-dependent retarded Greens function matrix $\hat{G}^\mathrm{r}$ is obtained from the differential equation
\begin{equation}
   \left[ \mathrm{i} \frac{\mathrm{d}}{\mathrm{d}t} + \omega + \mathrm{i} \Gamma - {\hat H}(t)\right] {\hat G}^\mathrm{r} \left(t, \omega \right) = {\hat 1} \; .
\end{equation}

\end{document}